\documentclass[useAMS,usenatbib]{mn2e}

\usepackage{epsfig}
\usepackage{longtable}
\usepackage{times} 
\newcommand{\sw}{$Swift$}

\def \src16465 {\mbox{IGR~J16465--4507}}
\def \IGRJ16479 {\mbox{IGR~J16479$-$4514}}

\def \inte {{$INTEGRAL$}}
\def \sw {{\em Swift}}
\def \ergsec{\hbox{erg s$^{-1}$}}
\def \ferg {erg cm$^{-2}$ s$^{-1}$}
\def \ATel {Astron.\ Tel.}
\def \apj {ApJ}
\def \apjs {ApJS}\def \aap {A\&A}
\def \mnras {MNRAS}

\title[Orbital period in IGR~J16465-4507]
{Detection of an orbital period in the 
supergiant high mass X-ray binary IGR~J16465$-$4507 with \emph{Swift}-BAT}

\author[V.\ La Parola et al.]{V.\ La Parola$^{1}$, G.\ Cusumano$^{1}$, P.\ Romano$^{1}$, A.\ Segreto$^{1}$, 
S.\ Vercellone$^{1}$, G. Chincarini$^{2,3}$  \\
$^{1}$INAF, Istituto di Astrofisica Spaziale e Fisica Cosmica,
        Via U.\ La Malfa 153, I-90146 Palermo, Italy\\
$^{2}$ INAF-Osservatorio Astronomico di Brera, I-23807 Merate (LC), Italy\\
$^{3}$ Universita degli Studi di Milano, Bicocca, I-20126 Milano, Italy\\
}

\begin{document}

\date{}

\pagerange{\pageref{firstpage}--\pageref{lastpage}} \pubyear{2010}

\maketitle

\label{firstpage}

\begin{abstract}
We analysed the IGR~J16465--4507 Burst Alert Teelescope survey data 
collected during the first 54 months of the \sw\ mission.
The source is in a crowded field and it is revealed through an ad hoc 
imaging analysis at a significance 
level of $\sim$14 standard deviations. The 15--50 keV average flux is 
$\sim3 \times 10^{-11}$ \ferg.
The timing analysis  reveals an orbital period of 
30.243$\pm$0.035 days. The folded light curve shows the presence of 
a wide phase interval of minimum intensity, lasting $\sim 20\%$ of the orbital 
period. This could be explained with a full 
eclipse of the compact object in an extremely eccentric orbit or with 
the passage of the compact source through a lower density wind at the orbit 
apastron.
The modest dynamical range  observed during the BAT monitoring  
suggests that IGR~J16465$-$4507 is a wind-fed system, continuously accreting 
from a rather homogeneous wind, and not a member of the Supergiant Fast X-ray
Transient class. 

\end{abstract}

\begin{keywords}
X-rays: binaries -- X-rays: individual: IGR~J16465$-$4507. 

\noindent
Facility: {\it Swift}

\end{keywords}


	\section{Introduction\label{sfxt7:intro}}

\begin{figure}
\begin{center}
\centerline{\includegraphics[width=7.5cm,angle=0]{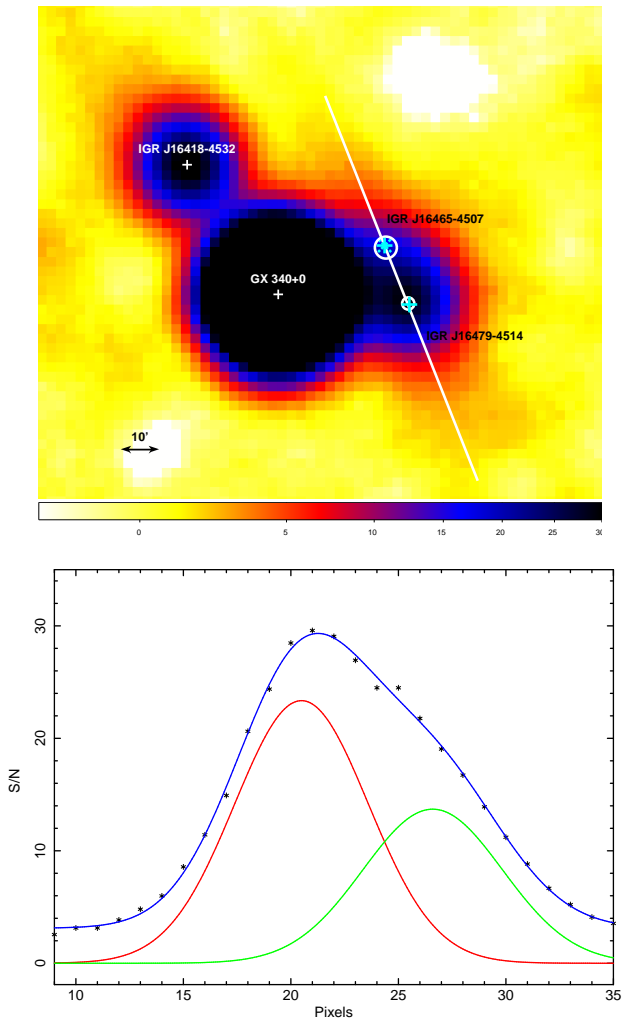}}
\centerline{\includegraphics[width=6cm,angle=270]{figure1b.ps}}
\caption[IGRJ16465-450 sky map]{{\bf Top}: 15--50 keV significance map in the 
neighborhood of IGR~J16465$-$4507. The color-bar represents the significance levels.
The cyan star marks the XMM position \citep{ZuritaHeras2004:16465-4507} of 
IGR J16465$-$4507.
The white circles are centered on the position
derived by fitting the significance profile extracted along the white line
with two Gaussians plus a constant; their radius corresponds to the 
90\% error on the position.
{\bf Bottom}: Significance profile extracted along the white line in the
above map. The plot shows the data (stars) and the best fit model (blue line, 
the sum of two Gaussian profiles plus a constant value). The higher peak (red line)
corresponds to IGR J16479$-$4514, the lower peak (green line) corresponds to 
IGR J16465$-$4507.
                }
		\label{map} 
        \end{center}
        \end{figure}

High mass X--ray binaries (HMXBs), stellar systems composed of a 
compact object and an early-type massive star, are traditionally divided 
in two subclasses \citep[e.g.][and references therein]{vanParadijs1995:binaries}, 
depending on the nature of the high mass primary and,
consequently, the different mass-transfer and accretion mechanism.  
On one side are the systems with main sequence Be primaries (Be-HMXBs).
They are generally wide ($P_{\rm orb}\ga 10$\,d)  
eccentric (eccentricity $e\sim 0.3$--0.5) systems in which the primaries are not filling 
their Roche lobe, and accretion onto the compact object occurs from the 
equatorial region of the rapidly rotating Be star.  
Most of these systems are highly variable: in some of them recurrent outbursts 
are observed caused by an enhanced rate when the compact star passes close to
the Be star. On the other side are the systems with an evolved OB supergiant primary (sgHMXB).
Their periods are shorter ($P_{\rm orb}\la 10$\,d) and their orbits more circular 
than in Be-HMXBs. They are powered either by a geometrically thin accretion disc 
or by the strong radiation-driven stellar wind, depending on whether 
the primary fills its Roche lobe or not. Their X-ray emission 
is bright and persistent. 

Recently, this rather clear-cut picture was made more structured with the INTEGRAL observations
of the Galactic plane. Two additional classes  were added to the classical OB
primary HMXBs: 
the highly absorbed persistent systems (Walter et al. 2004, 2006) and the supergiant fast X--ray transients
\citep[SFXTs, ][]{Sguera2005,negueruela06esa, smith04,intzand05}. 
The former are characterized by orbital and spin periods consistent with
those observed in wind-accreting systems, but a much higher absorbing column density.
The latter are transient sources showing a large dynamic range of 3--5 orders of magnitude with 
sporadic outbursts (which however are significantly shorter than those of typical Be-HMXBs), 
characterized by bright flares lasting
up to days with peak luminosities of 10$^{36}$--10$^{37}$~erg~s$^{-1}$ 
\citep{Sguera2005,Romano2009:sfxts_paperV,sidoli09}.

The Burst Alert Telescope \citep[BAT,][]{Barthelmy2005:BAT} on board \sw\ \citep{Gehrels2004mn}
is performing a continuous coverage of the hard X-ray sky (50 to 80\% of the sky every day).
This allowed the detection of many of the new INTEGRAL HMXBs (e.g. \citealp{cusumano10}) and the collection of their
long term light curves. 
In this Letter we analyse the hard X-ray data collected  during the
first 54 months of \sw-BAT sky monitoring
using data in the region of the IGR~J16465--4507.
This source was discovered by INTEGRAL in 2004 \citep{Lutovinov2004:16465-4507}  and 
X-ray activity was observed with IBIS/ISGRI starting on September 6, at 
a flux level of 8.8 $\pm$ 0.9 mCrab (18-60 keV), followed by a flare (up to
28 mCrab) on September 7. Follow-up observations with XMM-Newton revealed pulsations
at 228$\pm$6 s \citep{Lutovinov2005}  and allowed
the identification of the optical counterpart with 2MASS~J16463526--4507045
\citep{ZuritaHeras2004:16465-4507}.
This was classified as a B0.5 Ib supergiant at a distance of $\sim 8$\,kpc \citep{Negueruela2007} 
or as a O9.5 Ia supergiant at a distance of $9.5_{-5.7}^{+14.1}$\,kpc
\citep{Nespoli2008}.  
The supergiant nature of the companion, combined with the observed hard X-ray 
variability \citep{Lutovinov2004:16465-4507}  
and the X-ray spectral distribution, modeled
by a hard power law with photon index 1.0$\pm$0.52 \citep{Lutovinov2005},
suggested a classification of this source as a SFXT \citep{Negueruela2006}.
However, \citet{Walter2007}, based on  INTEGRAL measurements,  
suggested that  IGR~J16465--4507 is likely a classical supergiant 
HMXB, with an average flux just below the IBIS/ISGRI sensitivity undergoing 
sporadic long periods of enhanced activity.

This Letter is organized as follows. Section 2 describes the BAT data reduction and the
imaging analysis. Section 3 reports on the timing analysis. Sect. 4 describes the analysis of the 
pointed soft X-ray observation with \sw-XRT. In Sect.\ 5 we briefly discuss
our results. Errors are at 90\,\% confidence level, if not stated otherwise.

	\section{Imaging analysis\label{sfxt7:data}}

We analysed the BAT survey data of the first 54 months of the \sw\ 
mission. We retrieved the raw data from the HEASARC public 
archive\footnote{http://heasarc.gsfc.nasa.gov/cgi-bin/W3Browse/w3browse.pl} and
processed them with a dedicated software \citep{segreto10} that performs screening, 
mosaicking and source detection on data from coded mask instruments. The code 
also produces light curves for any given sky position. 

Figure~\ref{map} (top panel) shows the 15--50 keV significance sky map 
(exposure time of 17.7 Ms) in the direction of the sgHMXB IGR~J16465$-$4507.
The position of the source \citep{ZuritaHeras2004:16465-4507} is marked with
a cyan cross.
The source is in a crowded field, at 16.8 arcmin from the SFXT 
IGR~J16479$-$4514 and 30.8 arcmin from the low mass X-ray binary GX340+0 
and for this reason  it is not detected with our code. However, the
significance profile extracted along the line crossing the position of  
IGR~J16479$-$4514 and IGR~J16465$-$4507 shows an
asymmetric shape that can not be fitted with a constant plus a single Gaussian (the root mean square deviation
 is 48.8) and suggests the contribution of more than one source.
The profile is instead well fitted with a constant plus a double Gaussian (the root mean square deviation 
is 3.2 and the data points are uniformly distributed around the model)
peaking at the position of the two sources, with the weaker source (IGR~J16465$-$4507) at the 
significance of 13.7 standard deviations (Fig.~\ref{map}, bottom panel).

Moreover, the presence of IGR~J16465$-$4507 is confirmed by the analysis of the sky map produced using data 
selected in time intervals when the nearby source IGR~J16479$-$4514 was in eclipse 
\citep{Bozzo2008:eclipse16479,Jain2009:16479_period}. In this map
(3.4\,Ms exposure) IGR~J16465$-$4507 is detected with a signal-to-noise S/N$\sim8$.

\begin{figure}[h]
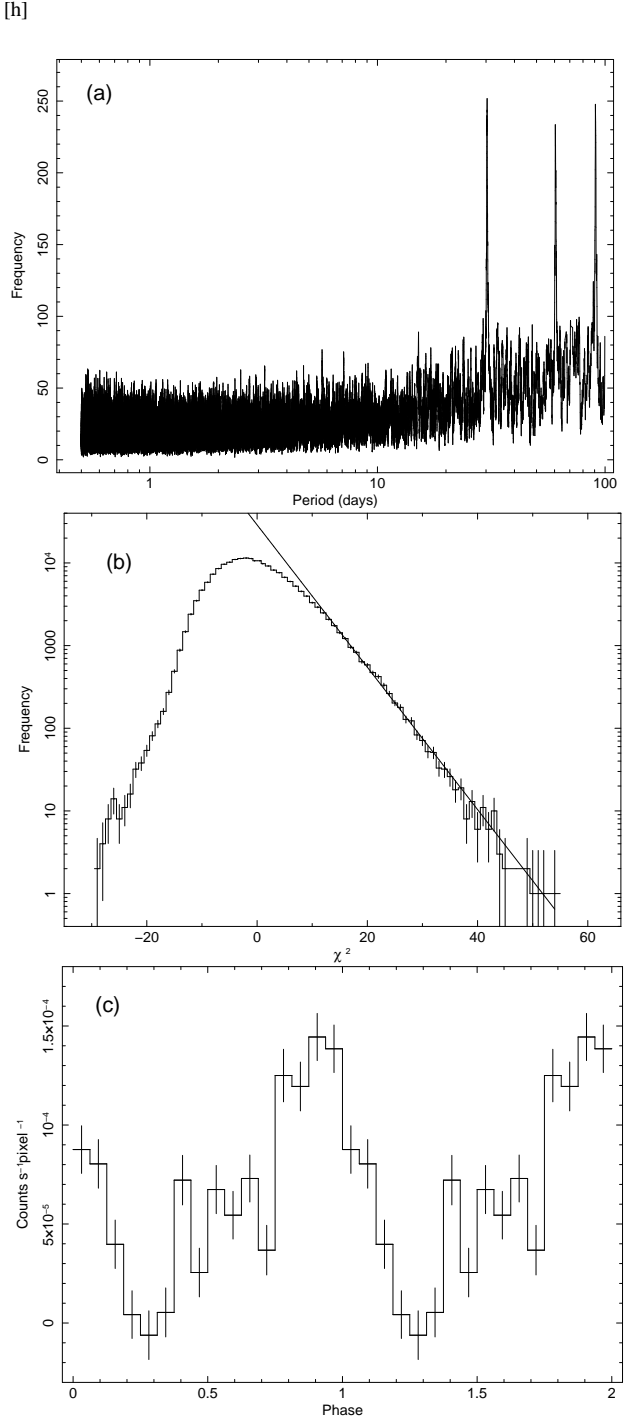

\begin{center}
\centerline{\includegraphics[width=6cm,angle=270]{figure2a.ps}}
\centerline{\includegraphics[width=6cm,angle=270]{figure2b.ps}}
\centerline{\includegraphics[width=6cm,angle=270]{figure2c.ps}}
\caption[]{{\bf (a)}: Periodogram of \sw-BAT (15--50\,keV) data for 
IGR~J16465--4507.
{\bf (b)}: Distribution of $\chi^2$ values extracted around 
$P_{\rm orb}$, in the period range between 0.5 and 40 days excluding the values between 
29.243 and 31.243 days. The continuous line is the best fit obtained with an
exponential model  applied to the tail of the distribution ($\chi^2>$ 15).
{\bf (c)}: \sw-BAT Light curve folded at a period $P=30.243$\,day, with 16 phase 
bins.
                }
		\label{period} 
        \end{center}
        \end{figure}


	\section{Timing analysis\label{sfxt7:timing}}

We extracted the light curve of IGR~J16465$-$4507 in the 15--50 keV energy range
with the maximum time resolution allowed by the data ($\sim300$\,s), subtracting the
contamination of the nearby sources as detailed in \citet{segreto10}.
The photon arrival times were corrected to the Solar system barycenter (SSB) by 
using the task {\sc earth2sun}. 

We looked for evidence of orbital periodicities in the BAT data.  
A folding technique was applied to the baricentered
arrival times by searching in the 0.5--100\,d period range with a step resolution 
of $P^{2}/(N \,\Delta T)$, where $N=16$ is the number of phase bins
 and $\Delta T$ (140,213,559.0 s) is the data time span. 
The average rate in each phase bin was evaluated by weighting the light curve rates 
by the inverse square of the corresponding statistical error:
\begin{equation}
R_j=\frac{\sum{r_i/er_i^2}}{\sum{1/er_i^2}}   
\end{equation}
where $R_j$ is the average rate in the j-th phase bin (j=1,16) of the 
trial profile, $r_i$ are the rates of the light curve whose phase fall 
into the j-th phase bin and $er_i$ are the
corresponding statistical errors. The error on $R_j$ is
$(\sqrt{\sum{1/er_i^2}})^{-1}$. 
This procedure, adopted to deal with the large span in $er_i$, is
justified by the fact that the data are background dominated.
Figure~\ref{period} (a) shows the 
resulting periodogram. 
We find significant evidence for periodicity ($\chi^2\sim251.8$)
at a period of $P_{\rm orb}=30.243\pm0.035$\,days (the two features at higher
periods are multiple of $P_{\rm orb}$) where the error is the
period resolution. We observe that the average $\chi^2$ in the periodogram is far from the
average value expected for white noise $(N-1)$ and increases with
increasing trial periods. As a consequence, the $\chi^2$ statistics cannot 
be applied to evaluate the significance of the detected periodicity. In order to
have an estimate of the significance of the observed feature we performed the
following steps:\\
(1) we corrected the $\chi^2$ distribution by subtracting the trend modeled with
a 2nd order polinomial fit. The value at 
$P_{\rm orb}$ subtracted for the pedestal is 206.7.\\
(2) we computed the histogram of the resulting $\chi^2$ distribution between 0.5 and 40 days 
[Figure~\ref{period} (b)], excluding
the peak region (29.243--31.243 days).\\
(3) we fit the distribution for $\chi^2 > 15$  with an exponential 
function and evaluated the integral of the best-fit function beyond 206.7. 
This integral yields a number of chance occurencies due to noise of 
$2.03\times10^{-13}$ corresponding to a significance of the detected temporal
feature of $\sim$7.3 standard deviations in 
Gaussian statistics.

As an alternative method to evaluate the significance of the temporal feature 
we produced $2000$ light curves
performing a random distribution of the observed rates in the bin times of 
the original light curve.
A periodogram between 0.5 and 100 days [51673.0 trial periods with the 
resolution of $P^{2}/(N \,\Delta T)$] 
was calculated for each of them obtaining a maximum $\chi^2$  value of 91.5 in
the whole sample. 
This corresponds to a significance of the 
observed periodicity higher than 5.5 standard deviations.  

In order to exclude the presence of systematic features in the periodogram, 
we performed the same analysis on the light curve of the two nearby sources 
(\IGRJ16479 {} and GX~340+0). We found no evidence for any significant 
feature at $P_{\rm orb}$.

In Fig.~\ref{period} (c) we show the pulsed profile folded at  
$P_{\rm orb}$ with $T_{\rm epoch}=54172.4236$ MJD.
There is clear evidence for a phase consistent with null intensity, whose 
centroid, evaluated 
by fitting the data around the dip with a Gaussian function, is at phase $0.268\pm
0.015$, corresponding to MJD (54180.5$\pm0.5$) $\pm nP_{\rm orb}$ .

\begin{figure}
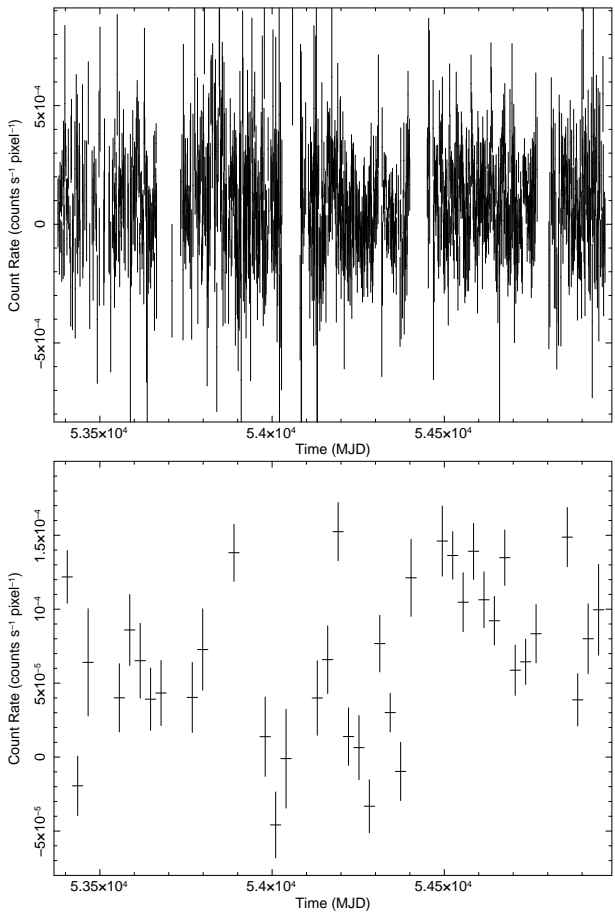

\begin{center}
\centerline{\includegraphics[width=6cm,angle=270]{figure3b.ps}}
\centerline{\includegraphics[width=6cm,angle=270]{figure3a.ps}}
\caption[]{{\bf Top panel}: BAT light curve with a bin time of 1 day .
{\bf Bottom panel}: BAT light curve with a bin time of $P_{\rm orb}$  .
                }
		\label{lc} 
        \end{center}
        \end{figure}

The BAT light curve at the time resolution of 1 day (Fig.~\ref{lc}) (top panel) shows no evidence for the 
presence of flare episodes throughout the period of monitoring. The
average count rate is $(6.9\pm0.5)\times 10^{-5}$ counts s$^{-1}$ pixel$^{-1}$ 
and the maximum count rate spread is 
consistent within 3 standard deviations with the average. 
Figure~\ref{lc} (bottom panel) shows the 54-months BAT light curve with a bin time of $P_{\rm orb}$, excluding bins 
with an exposure fraction less than 5\%. The source is detected in most of the time bins (30 out of 38). 
Adopting the best fit model of the INTEGRAL
data \citep{Lutovinov2005}, the average count rate corresponds to an observed
15--50 keV flux of $2.36\times 10^{-11}$ \ferg (2.3 mCrab). 
Assuming a distance of 9.5 kpc \citep{Nespoli2008}, the average 
luminosity is $2.6\times 10^{35}$ \ergsec.

\section{Soft X-ray data\label{sfxt7:xrt}}
\begin{figure}
\begin{center}
\centerline{\includegraphics[width=6cm,angle=270]{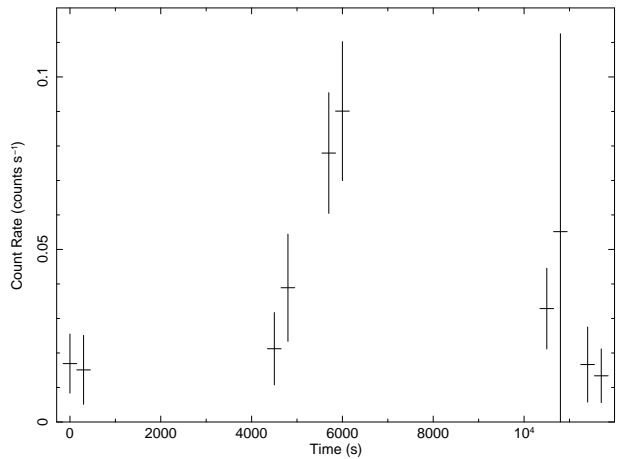}}
\caption[]{XRT light curve binned at 300 s.
                }
		\label{xrtlc} 
        \end{center}
        \end{figure}

%
The source was observed by \sw-XRT on 2009 June 13 (ObsId 00037885001, 
2.3\,ks net exposure). The epoch of the
observation corresponds to an orbital phase of $\sim0.2$. 
The data were processed with standard techniques with the 
{\sc ftools} in the {\sc Heasoft} package (v.6.8). 
Figure~\ref{xrtlc} shows the background subtracted 0.3--10 keV 
light curve, with a time bin of 300 s.

The statistic content of the XRT data is too low for a meaningful 
spectral analysis. In order to convert the observed
count rate into 0.3-10 keV flux we have used the best fit model
obtained by \citet{Lutovinov2005} for the XMM data. 
We obtain an observed flux in the 0.3--10\,keV energy band of 
$1.16\times 10^{-11}$ erg cm$^{-2}$ s$^{-1}$. 
This corresponds to an X--ray luminosity 
$L_{\rm X}\sim 1.2\times10^{35}$\,erg s$^{-1}$ in the same band.

\section{Discussion and conclusions\label{sfxt7:discuss}}

We have analysed the data collected by \sw-BAT during the first 54 months of 
the \sw\ mission in the region of the supergiant HMXB \src16465 {}.
The source is detected at a significance level of $\sim$ 14 standard deviations 
with an average flux of $\sim 2.36 \times 10^{-11}$ \ferg\ in the 15--50 keV 
energy band. 

The light curve reveals a periodicity of $30.243 \pm 0.035$ days that we 
interpret as  the orbital period of the binary system. 

%
%
By applying Kepler's third law, the semi-major axis of the binary system
is given by $a^3=P_{\rm orb}^2\times \rm G(M_{\star}+M_{\rm X})/4\pi^2$, 
where M$_{\star}$ and M$_{\rm X}$ are the masses of the supergiant and 
compact object, respectively. We adopt M$_{\rm X}=1.4$~M$_{\sun}$ and 
M$_{\star}=27.8$~M$_{\sun}$ 
\citep[][for an O9.5 I star of radius R$_{\star}=22.1$~R$_{\sun}$]{Martins2005}. 
This yields $a\sim125$~R$_{\sun} \sim 6$~R$_{\star}$.  
We note that the assumption of a stellar type B0.5 Ib would lead to an estimate of 
$a\sim150$~R$_{\sun} \sim 5$~R$_{\star}$ 
\citep[][M$_{\star}=47$~M$_{\sun}$, R$_{\star}=32.2$~R$_{\sun}$]{Searle2008}. 

The folded profile 
shows the presence of a dip with a count rate consistent with no emission, that could be
intepreted as a full eclipse. However, the width of this dip ($\sim 20\%$ of
P$_{\rm orb}$) is not consistent with the duration of the eclipse expected for a
circular orbit with 6 R$_{\star}$ semimajor axis, assuming an edge-on inclination. To obtain such a long eclipse, the system should have an
eccentricity of at least 0.8 coupled with a high inclination. This scenario is
unlikely. Alternatively, this wide phase interval of minimum intensity 
could be explained with an
eccentric orbit and a lower wind density at apastron.

\begin{figure}
\begin{center}
\centerline{\includegraphics[width=6cm,angle=90]{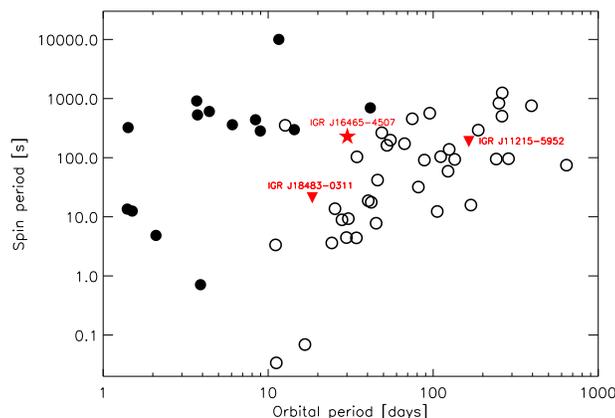}}
\caption[]{The Corbet diagram  
showing the neutron star $P_{\rm spin}$ vs.\ binary period $P_{\rm orb}$. 
Black circles are HMXBs with an OB primary, empty circles those with a Be one.  
Larger (red) symbols represent SFXTs: 
triangles are IGR~J18483$-$0311  and IGR~J11215$-$5952,
the star the newly determined position of \src16465 . 
 }
		\label{corbet} 
        \end{center}
        \end{figure}

Given our knowledge of both a spin period and an orbital period we can locate 
the source on the Corbet 
diagram \citep{Corbet1986:diagram} together with the known $P_{\rm spin}$
and $P_{\rm orb}$ of other binary systems \citep{Bildsten1997,Liu2006:hmxb}. 
We also plot in the diagram the
position of the two SFXTs IGR J18483-0311 \citep{Levine2006:igr18483,Zurita2009:sax1818.6_period} and 
IGR~J11215-5952 \citep{Swank2007:atel999,Sidoli2007,Romano2009:11215_2008}. \src16465 {} 
sits at the boundary of the wind-fed OB-HMXBs (systems with OB supergiants that underfill their 
Roche-lobes) and the locus of the Be transients (Fig.~\ref{corbet}. 

The characteristics of the system \citep{Lutovinov2004:16465-4507,Lutovinov2005}
suggested 
a classification of this source as a SFXT \citep{Negueruela2006}. However, 
the outburst history of this source is rather scarce.
After the initial discovery by \inte\ no further outbursts have been reported.
\citet{Walter2007} mention three episodes of enhanced hard X-ray activity 
observed by \inte\ at the limit 
of the instrumental sensitivity, that cannot be considered as real flares.
The BAT 15--50\,keV light curve during the 54 months of monitoring indicates 
that this is a  faint persistent source, with an average 
luminosity of $2.6\times10^{35}$ erg s$^{-1}$ (at a distance of 9.5 kpc), no 
evidence for flaring activity and a weak variability with a
dynamical range lower than 10.   Based on their X-ray variability, sgHMXB can 
be classified as classical and 
absorbed systems ( variability factor $<$20) or as SFXT (variability factor $>100$). The timing behavior 
of IGR~J16465$-$4507 suggests a wind-fed system, with a neutron star continuously accreting from a rather 
homogeneous wind. The luminosity, lower than what
observed in classical sgHMXB ($10^{36}-10^{37}$ erg s$^{-1}$), can be explained with the larger orbital
separation ($\sim$5 R$_{\star}$ compared to $\sim$2 R$_{\star}$ in classical 
systems).

\section*{Acknowledgments}

We thank the anonymous referee for suggestions that helped improve the paper.
This work was supported by contracts ASI I/011/07/0 and  I/088/06/0.


\bsp

\label{lastpage}

\end{document}